\documentclass[aps,twocolumn,pre,showpacs,superscriptaddress,floatfix]{revtex4}
\usepackage [dvips]{graphicx}
\usepackage {amssymb,amsmath}

\begin{document}

\title{Phase transition in the Ising model on a small-world network\\ 
with distance-dependent interactions}
\author{Daun \surname{Jeong}}
\affiliation{Department of Physics, Seoul National University,
Seoul 151-747, Korea}
\author{ H. \surname{Hong}}
\affiliation{School of Physics, Korea Institute for Advanced Study, Seoul 130-012, Korea}
\author{Beom Jun \surname{Kim}}
\affiliation{Department of Molecular Science and Technology, Ajou University, 
Suwon 442-749, Korea}
\author{M.Y. \surname{Choi}}
\affiliation{Department of Physics, Seoul National University,
Seoul 151-747, Korea}
\affiliation{Korea Institute for Advanced Study, Seoul 130-012, Korea}

\begin{abstract}
We study the collective behavior of an Ising system on a small-world 
network with the interaction $J(r) \propto r^{-\alpha}$, 
where $r$ represents the Euclidean distance between two nodes. 
In the case of $\alpha = 0$ corresponding to the uniform interaction, 
the system is known to possess a phase transition of the mean-field nature, 
while the system with the short-range interaction $(\alpha\rightarrow\infty)$ does not 
exhibit long-range order at any finite temperature.  
Monte Carlo simulations are performed at various values  of $\alpha$, and
the critical value $\alpha_c$ beyond which the long-range order does not
emerge is estimated to be zero.  
Thus concluded is the absence of a phase transition in the system with the 
algebraically decaying interaction $r^{-\alpha}$ 
for any nonzero positive value of $\alpha$.
\end{abstract}


\maketitle

Small-world networks, which are intermediate between the local regular networks
and the random networks, have two interesting features: high clustering, 
characteristic of regular networks, and short path length,
typically observed in random networks~\cite{ref:network,ref:WS}. 
The characteristic path length $l$ is defined to be the average of the shortest
path lengths between any two nodes: In many complex networks, the behavior
$l\sim$ log$N$ with the network size $N$ is observed, while 
regular lattices display $l\sim O(N)$. 
When a dynamical system of many elements 
is put on a small-world network, 
information exchange between two elements involves only
relatively short distance of the order of $\log N$. 
It has been observed in recent studies that the presence of shortcuts 
induces long-range order below the critical temperature~\cite{ref:Ising,ref:XY}. 
Furthermore, nature of the phase transition has been unambiguously demonstrated
to belong to the mean-field universality class. 
In those studies of the spin models on small-world networks, 
spin interactions have been assumed to be uniform, 
which is rather unrealistic in view of usual systems in nature. 
It is thus desirable to examine the general case
of distance-dependent interactions, which has not been addressed. 

In this paper we study the Ising model on a small-world network,
constructed from a one-dimensional lattice,
with the interaction $J$ decaying algebraically with distance $r$, 
i.e., $J(r) \propto r^{-\alpha}$. 
In such a case of the algebraically decreasing interaction,
the critical behavior is expected to change with the exponent $\alpha$: 
The limit $\alpha \rightarrow 0$ corresponds to the uniform interaction, 
which yields a mean-field transition at the finite critical 
temperature~\cite{ref:Ising}.  In the opposite limit of 
$\alpha \rightarrow \infty$, the spins interact with only 
their nearest neighbors, resulting in the absence of long-range order
at finite temperatures. 
This consideration suggests the existence of nontrivial critical value $\alpha_c$
beyond which the finite-temperature phase transition disappears. 

Here we construct a small-world network in the following way:
First a one-dimensional lattice of $N$ nodes is constructed with each node 
connected to its $2k$ nearest neighbors, where $k$ is the local interaction range.  
Then each local edge is visited once and a random long-range connection 
(shortcut) is added with the  probability $P$ without removing local edges. 
Note the difference from the original Watts and Strogatz (WS) 
construction~\cite{ref:WS}, where local edges are removed and reconnected to 
a randomly chosen node. 

The Hamiltonian for an Ising model on such a small-world network is given by
\begin{equation}
H=-\frac{1}{2}\sum_{i \neq j}J_{ij}\sigma_{i}\sigma_{j}, 
\end{equation}
where $\sigma_i \,(=\pm 1)$ is the Ising spin on node $i$ of the network. 
The distance-dependent interaction $J_{ij}$ reads
\begin{equation}
J_{ij}=J_{ji}=\left\{
\begin{array}{ll}
Jr_{ij}^{-\alpha}  & \textrm {if i and j are connected}\\
   0               & \textrm {otherwise,}
\end{array}
\right.
\end{equation}
where $r_{ij}$ is the geometrical distance (rather than
the shortest path length) between the two nodes $i$ and $j$
on the underlying one-dimensional lattice. 

From the comparison of the two limits $\alpha \rightarrow 0$ and 
$\alpha \rightarrow \infty$, a total of three distinct regimes may be 
expected~\cite{ref:Fisher,ref:Luijten}: 
(i) mean-field-type critical behavior for $\alpha<\alpha_1$;
(ii) continuously varying critical exponents for $\alpha_1 < \alpha < \alpha_2$;
(iii) the short-range interaction regime with no phase transition
for $\alpha>\alpha_2$. 
In Ref.~\onlinecite{ref:Fisher}, 
two boundary values $\alpha_1$ and $\alpha_2$ were obtained via 
renormalization-group calculations for the $O(n)$ model on a $d$-dimensional 
globally-coupled system. 
In particular, it is well known that $\alpha_1=3/2$ and $\alpha_2=2$
for the one-dimensional globally-coupled Ising model.  
Note here that the total number of connections in this one-dimensional 
globally-coupled system, given by $N(N-1)/2$ with $N$ being 
the number of spins, is far larger than that in a small-world network,
which is of the order of $N$. 
Accordingly, smaller boundary values of $\alpha$ are anticipated
for the small-world network. 

We have performed extensive Monte Carlo (MC) simulations with the heat bath algorithm 
at various values of $P$ and $\alpha$. 
The range $k=2$ was taken for convenience; other values of $k \,(>1)$ 
are not expected to give qualitatively different results. 
Measured in the simulations are Binder's cumulant~\cite{ref:Binder}, 
the susceptibility, and the specific heat: 
\begin{equation}
U_N=1-\frac{[\langle m^4\rangle]}{3[\langle m^2 \rangle]^2} 
\end{equation}
\begin{equation}
\chi = \frac{1}{N} \sum_{ij} [\langle \sigma_i \sigma_j \rangle]
\end{equation}
\begin{equation}
C_v = \frac{[\langle H^2 \rangle - {\langle H \rangle}^2]}{T^2 N}
\end{equation} 
with $m\equiv |(1/N)\sum_i \sigma_i|$. 
Here $\langle \cdots \rangle$ and $[ \cdots ]$ denote the thermal average, 
taken over $5\times 10^4$ MC steps after discarding $5\times 10^4$ MC steps 
for equilibration at each temperature, 
and the average over different network realizations, taken over 50 to 100 configurations, 
respectively. 
When $P=0$, the network reduces to a simple one-dimensional lattice, 
displaying no long-range order at any finite temperature. 
For $P \neq 0$, on the other hand, the presence of long-range shortcuts 
deprives the system of the locally connected one-dimensional character. 
In particular, when the interaction is uniform ($\alpha=0$), the system 
at $P \neq 0$ undergoes a finite-temperature transition of the 
mean-field nature~\cite{ref:Ising}. 

To find $\alpha_1$ and $\alpha_2$, beyond which mean-field behavior and 
long-range order do not emerge, respectively, 
we have examined the behavior of the system with $\alpha$ varied at fixed $P$. 
At any nonzero value of $\alpha$, the algebraically decaying interaction
is cut off by the finite network size and consequently, the ground-state 
energy in general decreases as $N$ grows. 
For given values of $k$, $P$, and $\alpha$, we thus normalize the coupling strength
$J$ so that the system has the size-independent ground-state energy per spin. 
As $\alpha$ is reduced, the interaction does not yet decay 
substantially at the boundary; this makes it inevitable
to study systems of very large sizes for obtaining correct scaling behavior. 

\begin{figure}
\includegraphics[width=8cm]{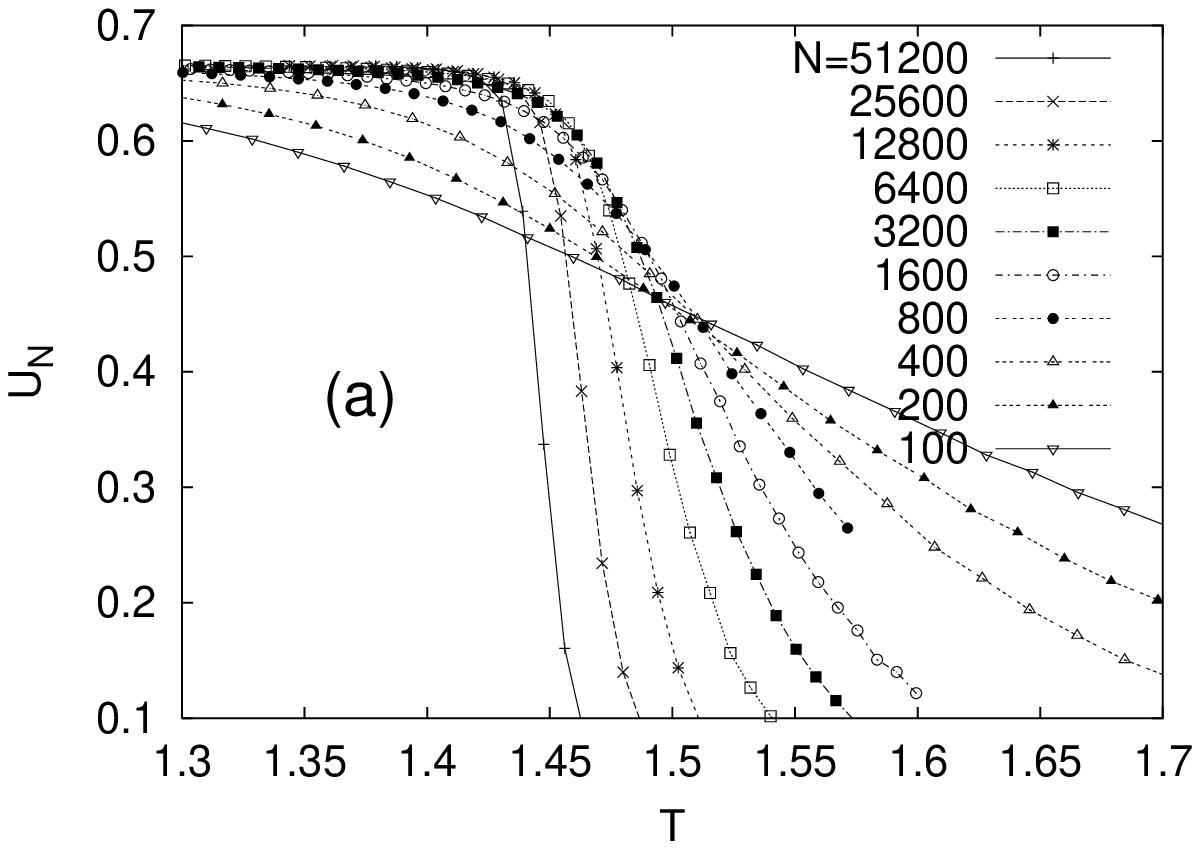}
\includegraphics[width=8cm]{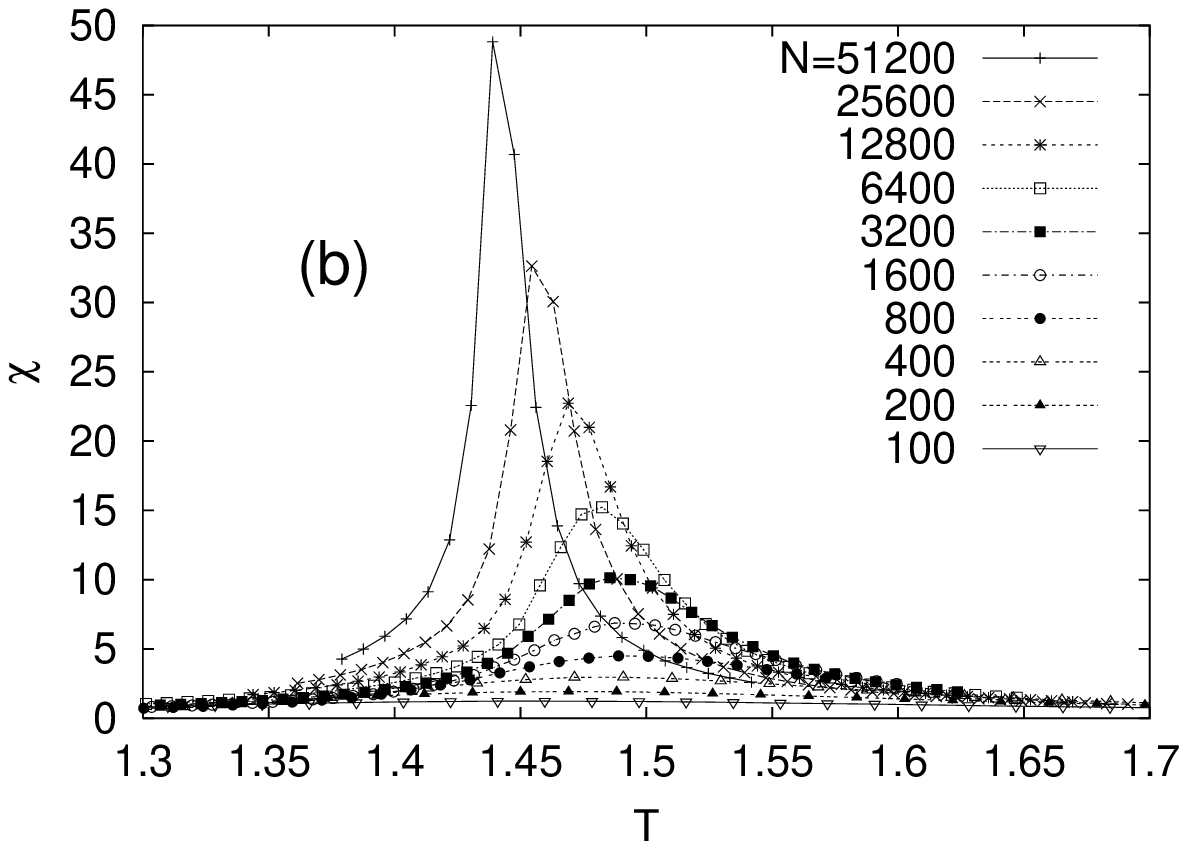}
\includegraphics[width=8cm]{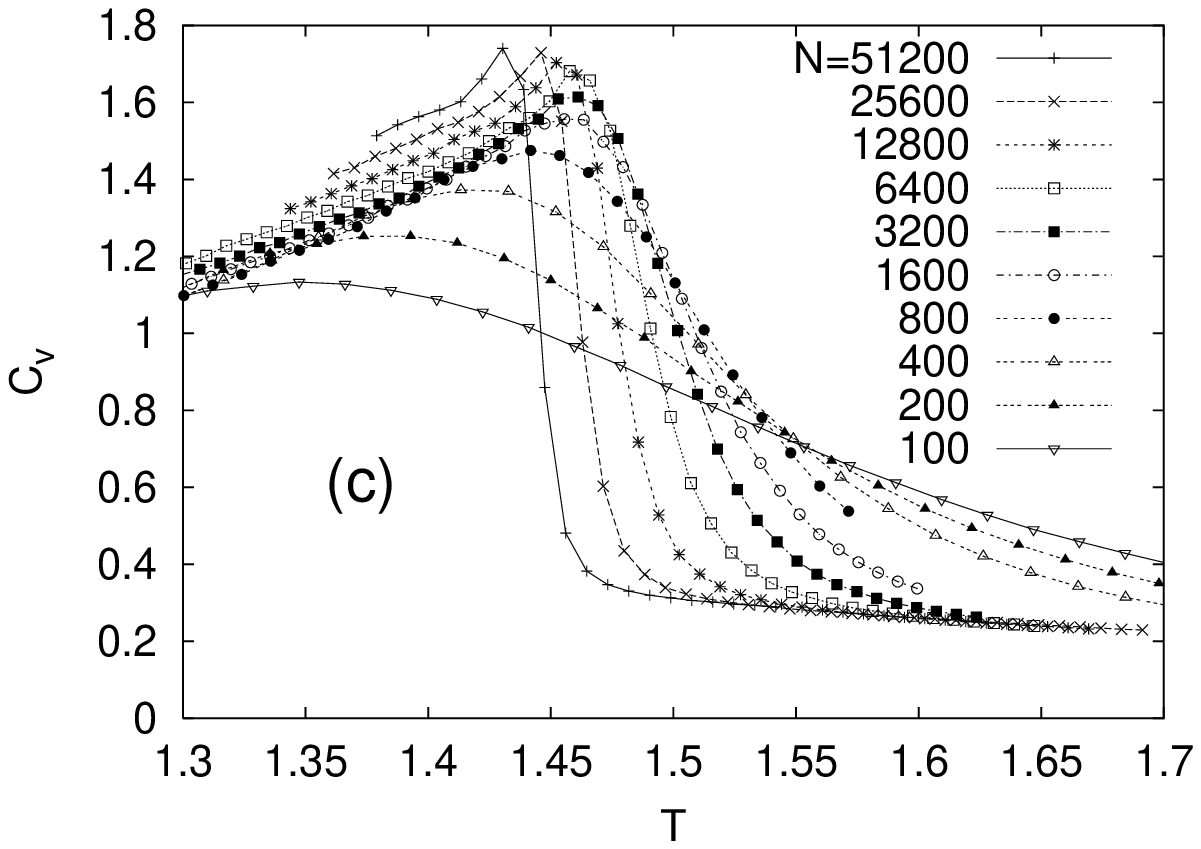}
\caption{Ising model on the small-world network with interaction
 $J_{ij}\propto r_{ij}^{-\alpha}$ for $P=0.5$, $k=2$ and $\alpha=0.1$. 
(a) Binder's cumulant $U_N$ possesses no unique crossing point.
(b) The peak temperature of the susceptibility $\chi$ keeps decreasing
for large system sizes ($N \gtrsim 1600$). 
(c) The specific heat $C_v$ does not display a unique crossing point; 
furthermore, peaks appear at lower temperatures as the system size 
gets large ($N\gtrsim 3200$). 
}
\label{fig:P0.5}
\end{figure}

Figure~\ref{fig:P0.5} presents the MC results for the network with $P=0.5$ 
and $\alpha=0.1$. 
Binder's cumulant $U_N$, the susceptibility $\chi$, and the specific heat $C_{V}$
as functions of the temperature are shown in (a), (b), and (c), respectively.
The absence of a unique crossing point of Binder's cumulant shown in 
Fig.~\ref{fig:P0.5}(a) and the shift of the susceptibility peak in Fig.~\ref{fig:P0.5}(b) 
as well as the peak position of the specific heat in Fig.~\ref{fig:P0.5}(c) 
unanimously suggest the absence of a phase transition at finite temperatures. 
At the temperatures where $\chi$ and $C_v$ display peaks, 
fluctuations are large and the correlation volume becomes comparable with the 
system size.  This gives some information 
about the temperature at which critical phenomena are observed. 
Figure~\ref{fig:P0.5}(b) shows that as the system size is increased,
the peak temperature of $\chi$ increases first ($N \leq 800$), in accord
with the observation in Ref.~\onlinecite{ref:Ising}, 
and begins to decrease eventually beyond $N=1600$. 
In particular the peak temperature does not saturate to a certain 
value but decreases continuously, supporting 
the absence of a finite-temperature transition. 
The specific heat exhibits similar behavior in Fig.~\ref{fig:P0.5}(c), 
although the system size at which the peak position starts to shift toward
lower temperatures does not coincide with that for the susceptibility. 

\begin{figure}
\includegraphics[width=8cm]{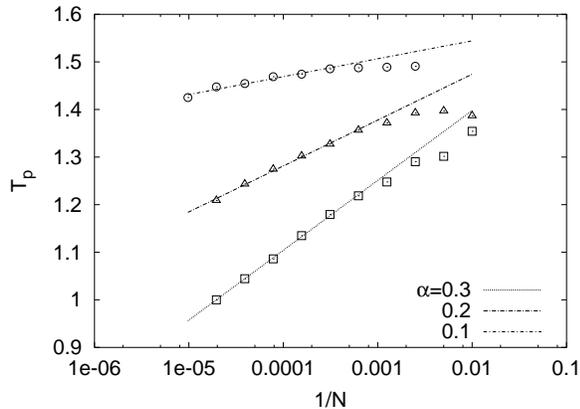}
\caption{Logarithmic behavior of the susceptibility peak temperature 
[see Fig.~\ref{fig:P0.5}(b)] with the inverse system size. 
Open circles, triangles, and squares represent data points for $P=0.5$ and 
$\alpha=0.1,\,0.2$, and 0.3, respectively, shown with the fitted lines. 
The larger the value of $\alpha$ is, the faster the peak temperature decreases. }
\label{fig:Tp}
\end{figure}

\begin{figure}
\includegraphics[width=8cm]{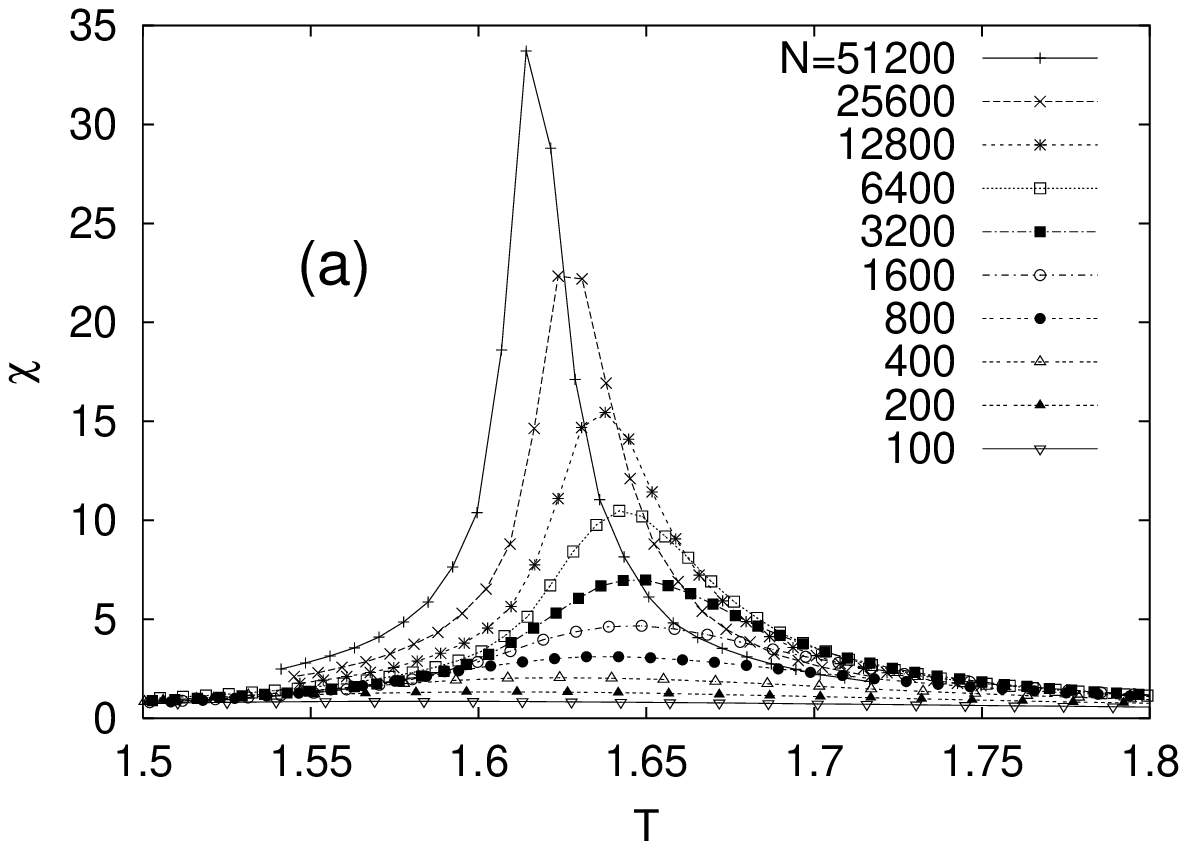}
\includegraphics[width=8cm]{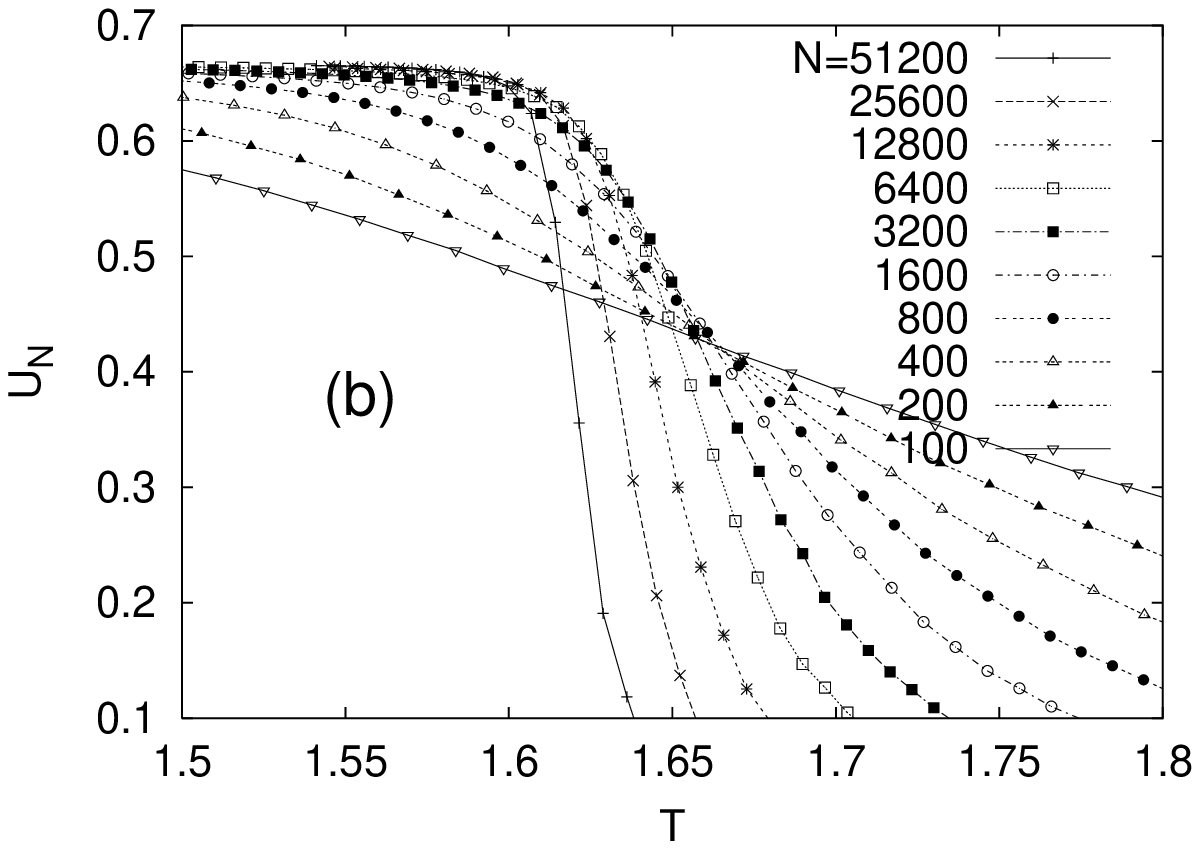}
\caption{Susceptibility $\chi$ and Binder's cumulant $U_N$ in the Ising model 
on a small-world network with interaction 
$J_{ij} \propto r_{ij}^{-\alpha}$ for $P=1.0$ and $\alpha=0.1$. 
(a) The peak temperature of the susceptibility decreases as the system size grows
for $N \gtrsim 1600$. (b) A unique crossing point is not observed in Binder's cumulant,
implying the absence of a phase transition at a finite temperature.}
\label{fig:P1.0}
\end{figure}

To clarify the drift of the peak temperature, 
we show in Fig.~\ref{fig:Tp} the peak temperatures $T_p$ of the susceptibility 
versus the inverse system size $1/N$, together with the fitted lines described by 
the logarithmic form: $T_p = a-b\log N$. 
The peak position thus approaches zero logarithmically as the system size grows. 
Here the values of the fitting parameters are given by $a=1.62$ and $b=0.016$ for 
$\alpha=0.1$; $a=1.67$ and $b=0.042$ for $\alpha=0.2$; $a=1.69$ and $b=0.064$ for 
$\alpha=0.3$. 
This indicates that the peak temperature decreases faster as the exponent $\alpha$ 
is raised.

As mentioned above, simulations here have been performed in the crossover region,
giving rise to logarithmic behavior of the peak position with the system size. 
When the exponent $\alpha$ is small, the interaction reduces very slowly with
the distance.  Nevertheless it gets weakened substantially in a system of very large size. 
For example, when $\alpha=0.1$, the interaction between the two connected 
nodes of the maximum distance (i.e., half the system size) is only about 40\% 
of the nearest-neighbor interaction in the system of size $N= 51200$. 
Consequently, long-range connections do not contribute significantly to the system
to be ordered in the thermodynamic limit, and the absence of a finite-temperature
transition is not very surprising. 
We have also investigated the case of $\alpha=0.05$, which requires simulations 
of even larger system size, to obtain similar results.  
These observations strongly suggest that the system does not undergo a phase transition 
at a finite temperature for any nonzero value of $\alpha$, 
thus leading to $\alpha_1=\alpha_2=0$. 
Figure~\ref{fig:P1.0} shows that the conclusion $\alpha_1 = \alpha_2 = 0$ 
is apparently valid even for $P=1.0$, which corresponds to 
the network with the largest number of long-range connections. 
Accordingly, the common belief that small-world networks are similar to 
globally-coupled networks in the transition nature~\cite{ref:Ising,ref:XY} 
is valid only for the (ideal) uniform interaction.  
In the more realistic case of the interaction
decaying with distance, the two systems display striking difference. 

In summary, we have studied the Ising model on small-world networks with 
the interaction decaying algebraically with exponent $\alpha$. 
For this, we have performed extensive Monte Carlo simulations on the networks
with the probability for adding shortcuts given by $P=0.5$ and $1$.
In both cases, absence of a finite-temperature transition has been observed
at any nonzero value of $\alpha$.
In particular, the absence in the latter, which is the extreme case of the small-world network 
with the largest number of long-range shortcuts, implies the validity for any values of $P\,(<1)$.
This conclusion should also hold for the original WS network, where
shortcuts are introduced by rewiring of the local edges,
since it has less interactions than the network investigated in 
the present work.

\acknowledgments
B.J.K. was  supported by the Korea Science and Engineering
Foundation through Grant No. R14-2002-062-01000-0.
M.Y.C. acknowledges the partial support from the Ministry of Education through the
BK21 Program. Numerical works have been performed on the computer
cluster Iceberg at Ajou University.

\end{document}